\documentclass{webofc}
\usepackage[varg]{txfonts}   
\usepackage{color}
\usepackage{bm}
\usepackage{epstopdf}
\usepackage{enumerate}
\usepackage{ulem}
\usepackage{hyperref}
\usepackage{physics}
\graphicspath{{./Figs/}}

\newcommand{\comment}[1]{}

\renewcommand\sout{\bgroup \color{red} \ULdepth=-.5ex \ULset}

\newcommand{\hyphen}{\mathchar`-}
\graphicspath{{./Figs/}}

\newcommand{\sk}{\rho}
\begin{document}
\renewcommand*{\thefootnote}{\fnsymbol{footnote}}
\title{
Directed flow of $\Lambda$ from heavy-ion collisions and hyperon puzzle of neutron stars\footnote{Report No.: YITP-22-122}}
%
%

\author{\firstname{Akira} \lastname{Ohnishi}\inst{1}\fnsep\thanks{\email{ohnishi@yukawa.kyoto-u.ac.jp}}
	\and
        \firstname{Asanosuke} \lastname{Jinno}\inst{2}
	\and
        \firstname{Koichi} \lastname{Murase}\inst{1}
	\and
        \firstname{Yasushi} \lastname{Nara}\inst{3}
}

\institute{
Yukawa Institute for Theoretical Physics, Kyoto University, Kyoto 606-8502, Japan
\and
Department of Physics, Faculty of Science, Kyoto University, Kyoto 606-8502, Japan
\and
Akita International University, Yuwa, Akita-city 010-1292, Japan
          }

\abstract{%
We examine the $\Lambda$ potential from the chiral effective field theory ($\chi$EFT)
via the $\Lambda$ directed flow from heavy-ion collisions.
We implement the $\Lambda$ potential obtained from the $\chi$EFT
in a vector potential version of relativistic quantum molecular dynamics.
We find that
the $\Lambda$ potentials obtained from the $\chi$EFT
assuming weak momentum dependence 
reproduce the $\Lambda$ directed flow measured by the STAR collaboration
in the Beam Energy Scan program.
While the $\Lambda$ directed flow is not very sensitive
to the density dependence of the potential,
the directed flow at large rapidities is susceptible
to the momentum dependence.
Thus understanding the directed flow of hyperons
in a wide range of beam energy and rapidity
is helpful in understanding hyperon potentials in dense matter.
}
\maketitle
\renewcommand*{\thefootnote}{\arabic{footnote}}
\section{Introduction}
\label{intro}
The hyperon puzzle is one of the central issues in neutron star physics
in view of hypernuclear and strange particle physics~\cite{HyperonPuzzle}.
When empirical two-body interactions are considered,
hyperons are predicted to appear
in the neutron star matter at $(2{\hyphen}4)\rho_0$,
soften the dense nuclear matter equation of state (EoS),
and reduce the maximum mass of neutron stars to be $M_\mathrm{max}=(1.3\hyphen1.6)M_\odot$.
Since several neutron stars are found to have masses around $2M_\odot$ or more,
most of the proposed hyperonic matter EoS were ruled out.
Thus something is wrong, and it is a challenge
to elucidate the mechanism to solve this hyperon puzzle.

Among several solution candidates to the hyperon puzzle,
we here concentrate on the solution based on the three-body $YNN$ repulsion~\cite{YNN}.
Since the three-body $NNN$ interactions are known to be necessary
to describe $A=3$ nuclear binding energies and saturation point of nuclear matter,
it is natural to expect that the three-body $YNN$ interactions
play some role in hypernuclei and hyperonic matter.
However, there is almost no information on three-body $YNN$ interactions 
from experiments, and we have to rely on theoretical predictions.
For example, the chiral effective field theory ($\chi$EFT) is
a first-principles framework and is expected to give reliable results
once the relevant low-energy constants are determined.
There are several predictions of the $\Lambda$ potential $U_\Lambda(\rho)$
at high densities obtained by using $\chi$EFT~\cite{GKW2020,Kohno2018}.
In Ref.~\cite{GKW2020}, Gerstung, Kaiser, and Weise (GKW) calculated $U_\Lambda$
in neutron star matter using $\chi$EFT
under the assumption that the diagrams relevant to the three-body forces
are saturated by the decuplet baryon propagation~\cite{Decuplet}.
The obtained $U_\Lambda$ with 2+3-body interactions
is more repulsive at high densities than that only with 2-body interactions,
and it suppresses $\Lambda$ to appear in neutron stars:
the single particle energy of $\Lambda$ at zero momentum is larger than the
neutron chemical potential, $M_\Lambda+U_\Lambda(\rho,\bm{p}=0) > \mu_n(\rho)$.
Confirmation of the repulsive $U_\Lambda$ at high densities is a challenge
in hypernuclear and strangeness particle physics.
Actually, one of the main goals of the J-PARC hadron hall extension~\cite{HEFex}
is to extract the three-body $YNN$ interaction via precision experiment
of hypernuclear spectroscopy.

Together with the hypernuclear spectroscopy,
hyperon observables from heavy-ion collisions should be useful,
since the formed nuclear matter reaches high densities.
Specifically, the anisotropic flows such as the directed flow,
$v_1=\langle \cos\phi \rangle$ with $\phi$ being the azimuthal angle
relative to the reaction plane, are generated in the initial stage
and have been utilized to constrain the equation of state (EoS).
Recently, the directed flow slopes ($dv_1/dy$)
for identified hadrons were measured in the colliding energy range of
$3~\mathrm{GeV}\leq \sqrt{s_{NN}} \leq 200~\mathrm{GeV}$~\cite{STARv1}.
The transition of the slope from positive to negative
was discovered for both protons and $\Lambda$s
at $\sqrt{s_{NN}}\approx10$ GeV\@.
Theoretical models with a first-order phase transition
predict this transition point at much lower beam energies%
~\cite{FirstOrder},
and the transition occurs at higher beam energies in hadron transport models
without potential effects~\cite{Konchakovski:2014gda}.
Recently, the proton directed-flow in the above colliding energy range
was explained by a transport model~\cite{JAMRQMDv} with a purely hadronic EoS\@.
A repulsive EoS contributes positively to the slope
in the early compression stage,
while the tilted ellipsoid of the formed matter causes a negative contribution
in the late expansion stage.
With increasing colliding energy,
the negative contribution takes over the positive one,
and then the above transition of the slope is realized.
The nuclear mean field including both the density and momentum dependence
explains the observed transition energy~\cite{JAMRQMDv}.

Thus, we expect that the $\Lambda$ directed flow is sensitive
to the $\Lambda$ potential at high densities.
Provided that the above-mentioned compression/expansion mechanism 
also applies to $\Lambda$,
one expects that the directed flow of $\Lambda$s should be small
since $\Lambda$s are produced during the compression stage
and the positive contribution would be smaller.
Then the agreement of the transition energies of protons and $\Lambda$s 
may suggest more repulsive potential for $\Lambda$s 
than for protons at higher densities, as suggested by the $\chi$EFT.

In the Hyp2022 presentation, we discussed the effects of $\Lambda$ potential
on the directed flow of $\Lambda$ by using an event generator JAM%
~\cite{JAM,JAMRQMDv}.
The density and momentum dependences of the $\Lambda$ potential
are taken from the $\chi$EFT
in Refs.~\cite{GKW2020} and \cite{Kohno2018}, respectively,
as discussed in Sec.~\ref{sec:lambdapot}.
In Sec.~\ref{sec:result}, we discuss the results of 
transport model calculations using JAM performed
in the relativistic quantum molecular dynamics mode
with a Lorentz-vector implementation of the potential (RQMDv)~\cite{JAMRQMDv}.
Readers are referred to Ref.~\cite{v1Lam}
for the details of the transport model and the calculated results.

\section{$\Lambda$ potential from chiral effective field theory}
\label{sec:lambdapot}

The $\Lambda$ potential as a function of baryon density has been studied extensively %
in Refs.~\cite{LambdaPot-NonRel,LambdaPot-Rel},
where the $\Lambda$ separation energies of various hypernuclei 
have been well fitted
and the EoS of the neutron-star matter has been predicted.
However, most of them fail to sustain massive neutron stars~\cite{HyperonPuzzle}.
While it is possible to increase the maximum mass of neutron stars above $2\,M_\odot$
by introducing repulsive three-body $YNN$ interactions
or repulsive hyperon potentials at high densities~\cite{YNN},
most of these prescriptions introduce additional parameters
which are not constrained by data or first principles.

\begin{figure}[tbhp]
\centering{
\includegraphics[width=6.5cm]{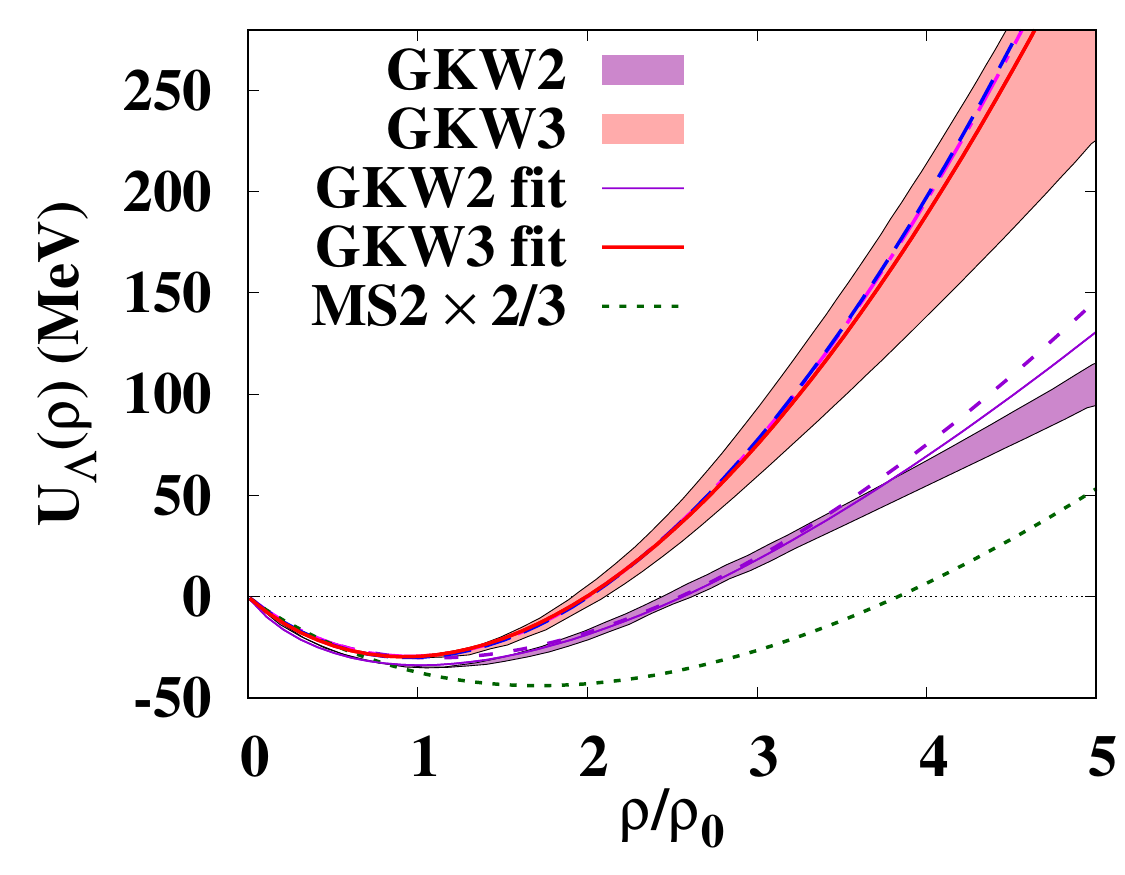}%
\includegraphics[width=6.5cm]{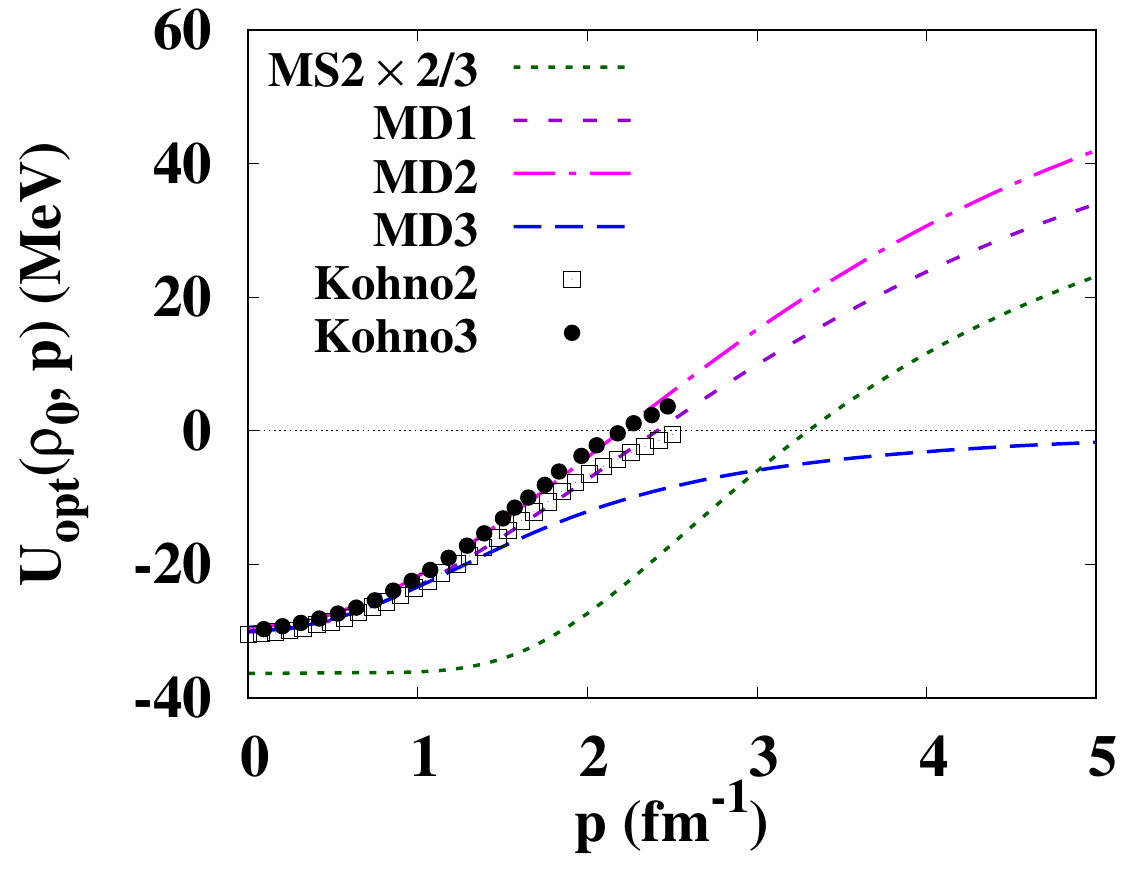}%
}
\caption{Baryon-density (left) and momentum (right) dependence
of the single-particle potentials for $\Lambda$.
GKW2 and GKW3 represent the $\Lambda$ single-particle potential from $\chi$EFT 
with 2- and 2+3-body interactions~\protect{\cite{GKW2020}}.
MS2$\times 2/3$ shows the nucleon single-particle potential
multiplied by $2/3$.
MD1, MD2, and MD3 are momentum-dependent potentials obtained by fitting the $\chi$EFT
results~\protect{\cite{Kohno2018}}.
}
\label{fig:pot}
\end{figure}

A promising way to systematically describe many-body interactions
is to invoke the $\chi$EFT, 
which is a first-principles theory based on the chiral symmetry of QCD.
As shown in the left panel of Fig.~\ref{fig:pot},
the $\Lambda$ potential in nuclear matter was computed by GKW~\cite{GKW2020}.
In uniform nuclear matter, 
the $\Lambda$ potential is given as a function of the density and momentum,
$U_\Lambda(\rho,\bm{p})$.
As shown in the right panel of Fig.~\ref{fig:pot},
the momentum dependence of $U_\Lambda$ in the $\chi$EFT is given,
for example, by Kohno~\cite{Kohno2018}.
By including the three-body $YNN$ interaction,
the $\Lambda$ potential becomes more repulsive at high densities and momenta.

We have parameterized the density and momentum dependence of the single-particle $\Lambda$ potential $U_\Lambda$ from $\chi$EFT in the Lorentz-vector type potential%
~\cite{Vector},
\begin{align}
U_\Lambda^\mu(\rho,\bm{p})
=&\frac{J^\mu}{\rho}U_{\sk\Lambda}(\rho)+U_{m\Lambda}^\mu(\rho,\bm{p})%
\label{eq:ULmu}\,,\\
U_{\sk\Lambda}(\rho)=&au+bu^{4/3}+cu^{5/3}
\label{eq:UL}\,,\\
U_{m\Lambda}^\mu (\rho,\bm{p})=& 
     \frac{C}{\rho_0}
     \int d^3{p}'
     \frac{p^{*'\mu}}{p^{*'0}}
     \frac{f(x,p')}{1+[(\bm{p}-\bm{p}')/\mu]^2},
     \label{eq:mdvec}
\end{align}
where
$J^\mu$ is the baryon current,
$u=\rho/\rho_0$ is the nucleon density normalized by the saturation density
$\rho_0=0.168$ fm$^{-3}$,
and $p^{*\mu} = p^\mu - U^\mu$
and $p^{*0}=\sqrt{m_N^2+\bm{p}^{*2}}$. 

In Fig.~\ref{fig:pot}, we show the fitting results of the $\Lambda$ potential.
MD1 and MD2 potentials (right panel) are obtained by
fitting the $\chi$EFT results by Kohno~\cite{Kohno2018}
at $\rho=\rho_0$
with 2- and 2+3-body interactions, respectively,
up to the momentum of 2.5 fm$^{-1} \simeq \lambda$ (cutoff)
assuming the range parameter $\mu=3.23$ fm$^{-1}$~\cite{JAMRQMDv}.
The real part of the optical potential $U_\mathrm{opt}$ is defined
as the difference between the single-particle energy and the kinetic energy.
Since the momentum dependence in Ref.~\cite{Kohno2018} seems to be stronger
than the previous estimate~\cite{fss2},
the potential with a weaker momentum dependence (MD3) is also prepared.
For MD3, we require that
the optical potential is zero at high momentum
($p\simeq8.6~\mathrm{fm}^{-1}$)
and reproduces the result of Ref.~\cite{Kohno2018}
at $p=1~\mathrm{fm}^{-1}\simeq 0.4\,\lambda$.
The density dependence (left panel) is determined by fitting the $\chi$EFT results
by GKW~\cite{GKW2020}.
GKW2 and GKW3 potentials are those with 2- and 2+3-body interactions,
respectively.
Since the Br\"uckner-Hartree-Fock calculation was found to be unstable
at $\rho/\rho_0>3.5$~\cite{GKW2020},
we fit the $\chi$EFT results in the range $\rho/\rho_0\leq3$.
At zero $\Lambda$ momentum in cold nuclear matter,
we find
$J^\mu=(\rho,\bm{0})$ and $\bm{U}_{m\Lambda}=\bm{0}$,
and thus
$U_\Lambda(\rho)\equiv U^0_\Lambda(\rho,\bm{0})=U_{\rho\Lambda}(\rho)+U^0_{m\Lambda}(\rho,\bm{0})$.
The density dependence from the momentum-dependent interaction (second term)
can be absorbed by tuning parameters $(a,b,c)$,
so the density dependence of $U_\Lambda$ is almost the same
with and without the momentum-dependent potentials:
GKW2 fit (thin solid, without momentum dependent terms) and GKW2+MD1 (dashed),
and 
GKW3 fit (thick solid) and GKW3+MD2/MD3 (dash-dotted, long dashed)
at $\rho/\rho_0\leq 3$.

\section{Directed flow of $\Lambda$ from heavy-ion collisions}
\label{sec:result}

In order to simulate high-energy heavy-ion collisions,
we use an event generator JAM2.1~\cite{jam2.1}.
In JAM2.1, there are several updates:
one can utilize Pythia~8~\cite{Pythia8} library as it is,
potentials for leading baryons are included during their formation
time with the reduced factor,
and collision time and ordering time has been modified~\cite{Zhao:2020yvf}.

We implement the $\Lambda$ potential described in the previous section
in the form of Lorentz-vector potential
in the relativistic quantum molecular dynamics (RQMDv) approach~\cite{JAMRQMDv}.
The RQMDv equations of motion for the $i$-th particle
having the position $q^\mu_i$ and the momentum $p^\mu_i$
is given by~\cite{RQMDv} 
\begin{align}
\frac{dq^\mu_i}{dt} =
   v^{*\mu}_i
     -\sum_{j} 
      v^{*\nu}_j
      \frac{\partial {V}_{j\nu}}{\partial p_{i\mu}}\,,\ 
\frac{dp^\mu_i}{dt}=
      \sum_{j} v^{*\nu}_j
     \frac{\partial V_{j\nu}}{\partial q_{i\mu}},
\end{align}
where $v_i^{*\mu}=p_i^{*\mu}/p_i^{*0}$.
For the momentum-dependent part, 
we replace $(\bm{p}_i-\bm{p}_j)^2$ in Eq.~\eqref{eq:mdvec}
with the two-body relative momentum squared, $p_{R,ij}^2$,
in the rest frame of the particle $j$.
We use $U_\Lambda^\mu$ in Eq.~\eqref{eq:ULmu} for $\Lambda$ and other hyperons.%
\footnote{In the HYP2022 presentation, I answered to the question by Hoai Le
that we do include different potentials for $\Sigma$ and $\Xi$,
but it was not correct.
The results shown in HYP2022 and in this proceedings
are obtained by assuming the same potential for all hyperons.}

\begin{figure}[tbhp]
\begin{center}
\includegraphics[width=6.5cm]{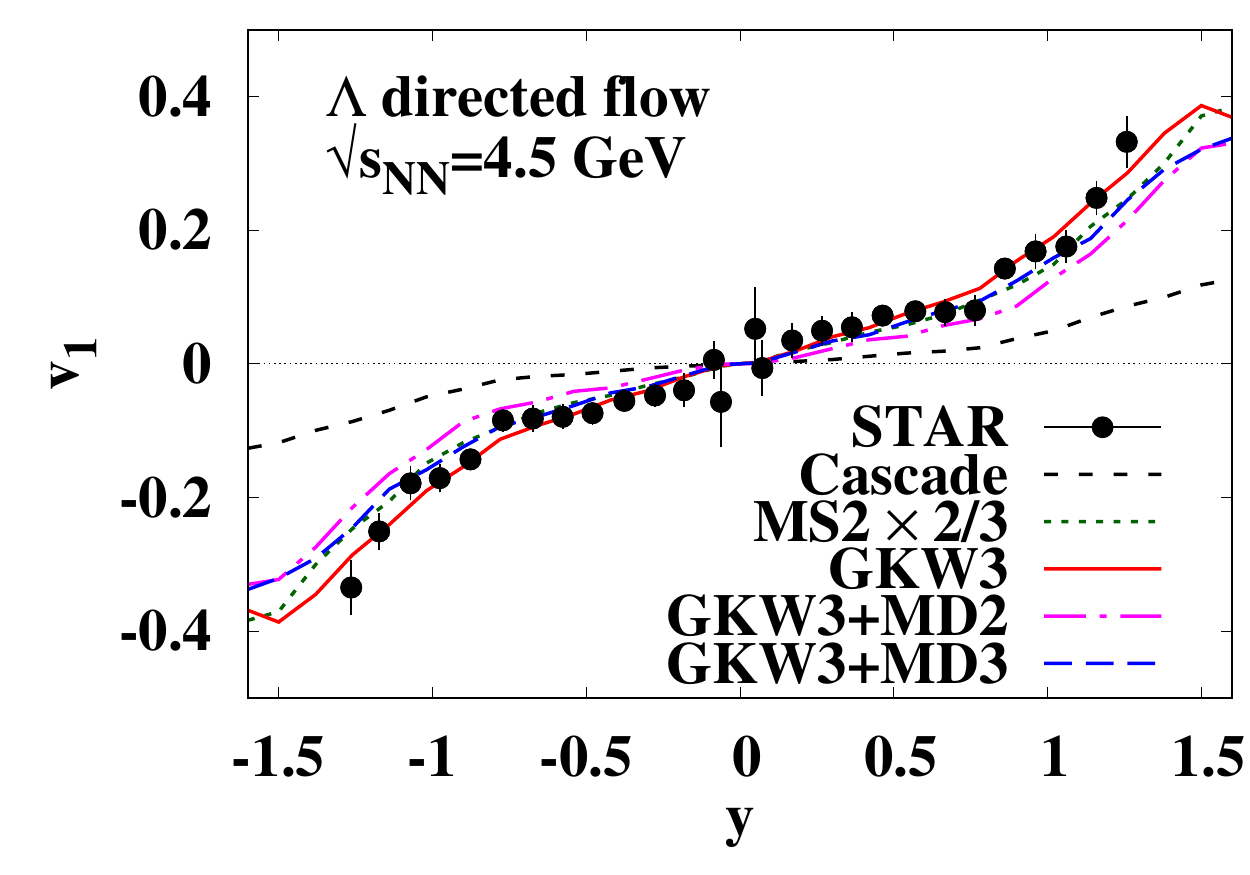}%
\includegraphics[width=6.5cm]{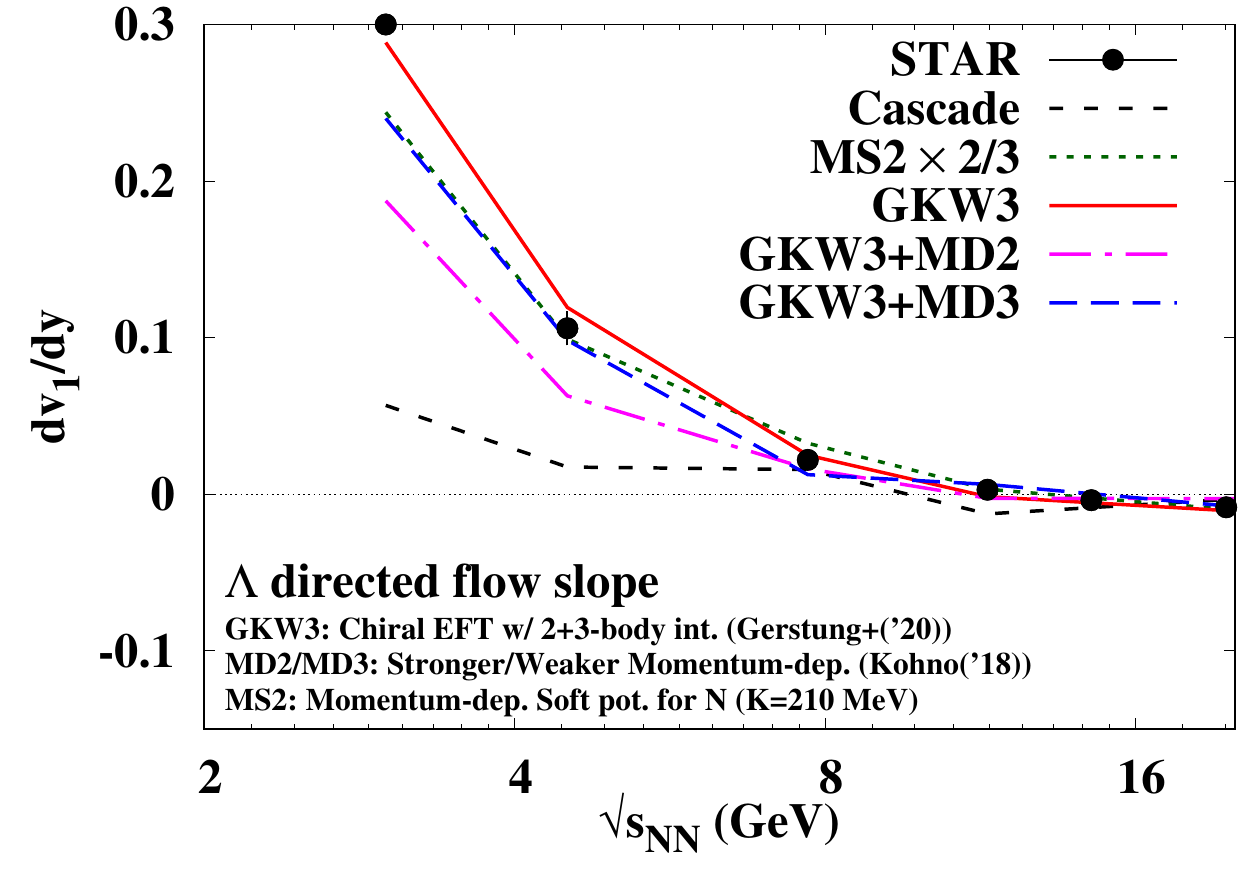}
\end{center}
\caption{Left: Calculated directed flow of $\Lambda$ in mid-central Au + Au
collisions at $\sqrt{s_{NN}}=4.5$ GeV.
Right: The directed flow slope of $\Lambda$.
Data are taken from Ref.~\cite{STARv1}.
}
\label{fig:v1lam}
\end{figure}

In Fig.~\ref{fig:v1lam},
we compare the calculated $\Lambda$ directed flow $v_1=\langle\cos\phi\rangle$
in the impact parameter range $4.6~\mathrm{fm}<b<9.4~\mathrm{fm}$
with the STAR data~\cite{STARv1} in midcentral Au + Au collisions
at $\sqrt{s_{NN}}=4.5$ GeV (left panel),
and the directed flow slope of $\Lambda$
at $\sqrt{s_{NN}}=(3.0\hyphen19.6)~\mathrm{GeV}$ (right panel). 
The cascade results (without potential effects) underestimate the data.
With the potential effects, the directed flow data of $\Lambda$ are roughly explained.
Specifically, GKW3 without momentum dependence explains the data well.\footnote{
In the HYP2022 presentation, we showed the calculated results
with an incorrect $p_T$ cut at $\sqrt{s_{NN}}=3.0~\mathrm{GeV}$,
and the GKW3 results underestimated the directed flow data.
By using the correct cut adopted in the experimental data analysis,
$0.4~\mathrm{GeV} \leq p_T \leq 2.0~\mathrm{GeV}$~\cite{STARv1},
the results from GKW3 explain the directed flow well
also at $\sqrt{s_{NN}}=3.0~\mathrm{GeV}$.
}

Now let us discuss the sensitivity of the directed flow 
to the density- and momentum-dependence of the $\Lambda$ potential.
Unfortunately, GKW2, not shown in the figure, gives the $\Lambda$ directed flow
similar to that from GKW3.
Thus we cannot discriminate the $\Lambda$ potentials from $\chi$EFT
with 2- and 2+3-body interactions by the directed flow.
We also note that the strong momentum dependence of the $\Lambda$ potential,
e.g. GKW3+MD2, reduces the directed flow, especially at large rapidities~\cite{v1Lam}.
We do not completely understand the origin of this dependence,
but we guess that the cancellation of the contributions
in the compression and expansion stages is one of the origins.
The repulsive potential at high densities increases the directed flow
in the compression stage, but a part of the increase is canceled by the negative
contribution from the expansion of the tilted matter.

\section{Summary and discussion}
\label{sec:summary}

We have investigated the $\Lambda$ potential via the $\Lambda$ directed flow
from heavy-ion collisions~\cite{v1Lam}.
We implement the $\Lambda$ potential from the $\chi$EFT~\cite{GKW2020,Kohno2018}
parameterized as a function of density and momentum
in the RQMDv mode of an event generator JAM.
The calculated results using the $\Lambda$ potential from $\chi$EFT
do reproduce the $\Lambda$ directed flow in a wide beam energy range.
This is the first support by experimental data
of the strongly repulsive $\Lambda$ potential at high densities.
It should be noted that similar results are obtained for the sideward flow
$\langle p_x \rangle$ at $\sqrt{s_{NN}}=2.7~\mathrm{GeV}$
in Ref.~\cite{Zhang:2021ddb}.

Yet there are several questions and problems.
First, the directed flow of $\Lambda$ can also be reproduced
by the $\Lambda$ potential only from the 2-body interactions
and with a weaker repulsion at high densities.
Next, the $\Lambda$ directed flow depends on the momentum dependence of the potential,
especially at large rapidities.
Provided that the compression/expansion mechanism~\cite{JAMRQMDv} is correct,
the directed flow of $\Lambda$s produced during the compression stage
is affected by the expansion of tilted matter
and would be more sensitive to dependence on momentum than density.
More analyses of the evolution are needed to understand 
the sensitivity to the details of the $\Lambda$ potential.

It is interesting to study $\Xi$ and $\Sigma$ flows as well.
These are related to
the repulsive $\Sigma$ potential in dense matter~\cite{Zhang:2021ddb},
the dense matter EoS~\cite{Yong:2021npa},
and
the formation of deconfined matter~\cite{Nayak:2019vtn}.

\medskip

The authors would like to thank Yue-Hang Leung for telling us
that the $p_T$ cut is different at $\sqrt{s_{NN}}=3.0~\mathrm{GeV}$.
This work was supported in part by the
Grants-in-Aid for Scientific Research from JSPS
(No. JP21K03577, 
No. JP19H01898, 
and No. JP21H00121
).

\end{document}